\documentclass[preprint,prd,tightenlines, superscriptaddress]{revtex4-1}

\usepackage{graphicx} 
\usepackage{dcolumn}  
\usepackage{colordvi}
\usepackage{color}
\usepackage{epstopdf}
\usepackage{pstricks}
\usepackage{amssymb}
\usepackage{url}
\graphicspath{{ps}}
\usepackage{hyperref}
\usepackage{tabularx}
\usepackage{multirow}
\usepackage{units}
\usepackage{upgreek}
\usepackage{siunitx}
\usepackage{hyphenat}
\usepackage{subfigure}
\usepackage[italic]{hepnames}

\newcommand{\CP}{\ensuremath{C\hspace{-0.13em}P}\xspace}

\newcolumntype{Y}{>{\centering\arraybackslash}X}

\def\mbc{\ensuremath{M_{{\rm bc}}}\xspace}

\def\ACP{{\ensuremath{\mathcal{A}_{\CP}}\xspace}}
\def\SCP{{\ensuremath{\mathcal{S}_{\CP}}\xspace}}

\RequirePackage{xspace}

\def\belletwo {Belle~II}

\def\en         {\ensuremath{e^-}\xspace}   
\def\ep         {\ensuremath{e^+}\xspace}

\def\qbar  {\ensuremath{\overline q}\xspace}
\def\qqbar {\ensuremath{q\overline q}\xspace}

\def\dbar  {\ensuremath{\overline d}\xspace}

\def\piz   {\ensuremath{\pi^0}\xspace}

\def\Kbar  {\kern 0.2em\overline{\kern -0.2em K}{}\xspace}

\def\Kz    {\ensuremath{K^0}\xspace}
\def\Kzb   {\ensuremath{\Kbar^0}\xspace}
\def\KzKzb {\ensuremath{\Kz \kern -0.16em \Kzb}\xspace}
\def\Kp    {\ensuremath{K^+}\xspace}
\def\Km    {\ensuremath{K^-}\xspace}

\def\KpKm  {\ensuremath{\Kp \kern -0.16em \Km}\xspace}
\def\KS    {\ensuremath{K^0_{\scriptscriptstyle S}}\xspace} 
\def\KL    {\ensuremath{K^0_{\scriptscriptstyle L}}\xspace}

\def\Dbar    {\kern 0.2em\overline{\kern -0.2em D}{}\xspace}

\def\Dz      {\ensuremath{D^0}\xspace}
\def\Dzb     {\ensuremath{\Dbar^0}\xspace}
\def\DzDzb   {\ensuremath{\Dz {\kern -0.16em \Dzb}}\xspace}
\def\Dp      {\ensuremath{D^+}\xspace}
\def\Dm      {\ensuremath{D^-}\xspace}

\def\DpDm    {\ensuremath{\Dp {\kern -0.16em \Dm}}\xspace}

\def\Bbar    {\kern 0.18em\overline{\kern -0.18em B}{}\xspace}

\def\BB      {\ensuremath{B\Bbar}\xspace} 
\def\Bz      {\ensuremath{B^0}\xspace}
\def\Bzb     {\ensuremath{\Bbar^0}\xspace}
\def\BzBzb   {\ensuremath{\Bz {\kern -0.16em \Bzb}}\xspace}
\def\Bu      {\ensuremath{B^+}\xspace}
\def\Bub     {\ensuremath{B^-}\xspace}

\def\BpBm    {\ensuremath{\Bu {\kern -0.16em \Bub}}\xspace}

\def\jpsi     {\ensuremath{{J\mskip -3mu/\mskip -2mu\psi\mskip 2mu}}\xspace}

\mathchardef\Upsilon="7107
\def\Y#1S{\ensuremath{\Upsilon{(#1S)}}\xspace}

\def\FourS {\Y4S}

\mathchardef\Deltares="7101
\mathchardef\Xi="7104
\mathchardef\Lambda="7103
\mathchardef\Sigma="7106
\mathchardef\Omega="710A

\def\Deltabar{\kern 0.25em\overline{\kern -0.25em \Deltares}{}\xspace}
\def\Lbar{\kern 0.2em\overline{\kern -0.2em\Lambda\kern 0.05em}\kern-0.05em{}\xspace}
\def\Sigbar{\kern 0.2em\overline{\kern -0.2em \Sigma}{}\xspace}
\def\Xibar{\kern 0.2em\overline{\kern -0.2em \Xi}{}\xspace}
\def\Obar{\kern 0.2em\overline{\kern -0.2em \Omega}{}\xspace}
\def\Nbar{\kern 0.2em\overline{\kern -0.2em N}{}\xspace}
\def\Xb{\kern 0.2em\overline{\kern -0.2em X}{}\xspace}

\newcommand{\tev}{\ensuremath{\mathrm{\,Te\kern -0.1em V}}\xspace}
\newcommand{\gev}{\ensuremath{\mathrm{\,Ge\kern -0.1em V}}\xspace}
\newcommand{\mev}{\ensuremath{\mathrm{\,Me\kern -0.1em V}}\xspace}
\newcommand{\kev}{\ensuremath{\mathrm{\,ke\kern -0.1em V}}\xspace}
\newcommand{\ev}{\ensuremath{\mathrm{\,e\kern -0.1em V}}\xspace}
\newcommand{\gevc}{\ensuremath{{\mathrm{\,Ge\kern -0.1em V\!/}c}}\xspace}
\newcommand{\mevc}{\ensuremath{{\mathrm{\,Me\kern -0.1em V\!/}c}}\xspace}
\newcommand{\gevcc}{\ensuremath{{\mathrm{\,Ge\kern -0.1em V\!/}c^2}}\xspace}
\newcommand{\mevcc}{\ensuremath{{\mathrm{\,Me\kern -0.1em V\!/}c^2}}\xspace}

\def\invfb   {\ensuremath{\mbox{\,fb}^{-1}}\xspace}
\def\invab   {\ensuremath{\mbox{\,ab}^{-1}}\xspace}

\def\mus  {\ensuremath{\rm \,\mus}\xspace}

\def\mus        {\ensuremath{\,\mu{\rm s}}\xspace}    

\def\to                 {\ensuremath{\rightarrow}\xspace}

\newcommand{\stat}{\ensuremath{\mathrm{(stat)}}\xspace}
\newcommand{\syst}{\ensuremath{\mathrm{(syst)}}\xspace}

\def\gsim{{~\raise.15em\hbox{$>$}\kern-.85em
          \lower.35em\hbox{$\sim$}~}\xspace}
\def\lsim{{~\raise.15em\hbox{$<$}\kern-.85em
          \lower.35em\hbox{$\sim$}~}\xspace}

\begin{document}

\def\belletwo {\it {Belle~II}}

\clubpenalty = 10000  
\widowpenalty = 10000 

\vspace*{-3\baselineskip}
\resizebox{!}{3cm}{\includegraphics{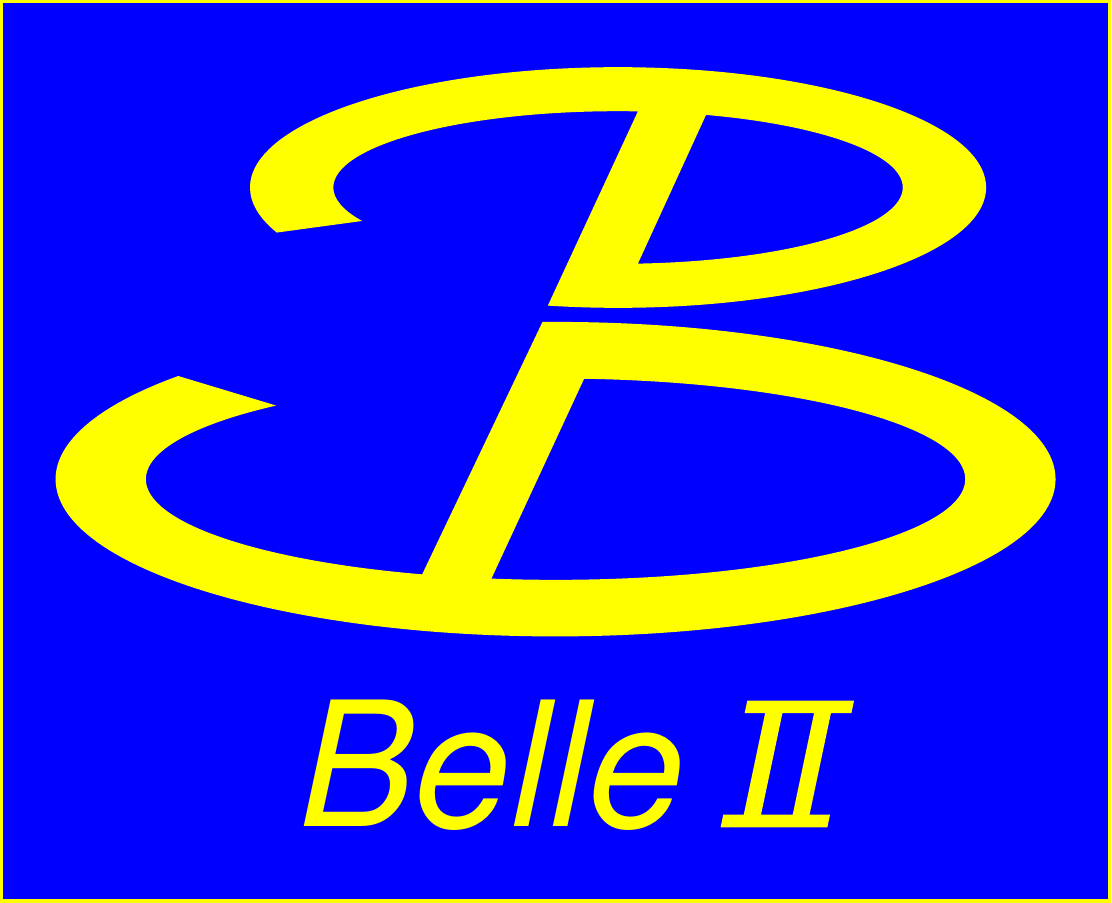}}

\vspace*{-5\baselineskip}
\begin{flushright}
BELLE2-CONF-2022-002

\today
\end{flushright}

\quad\\[0.5cm]

\title {
First decay-time-dependent analysis of $\Bz \to \KS \piz$ at Belle~II}



  \author{F.~Abudin{\'e}n}
  \author{I.~Adachi}
  \author{R.~Adak}
  \author{K.~Adamczyk}
  \author{L.~Aggarwal}
  \author{P.~Ahlburg}
  \author{H.~Ahmed}
  \author{J.~K.~Ahn}
  \author{H.~Aihara}
  \author{N.~Akopov}
% ...

\collaboration{Belle II Collaboration}

\begin{abstract}
We report measurements of the branching fraction ($\mathcal{B}$) and direct $\CP$-violating asymmetry ($\ACP$) of the charmless decay $\Bz\to K^0\pi^0$ at Belle~II. 
A sample of $\ep\en$ collisions, corresponding to $189.8\invfb$ of integrated luminosity, recorded at the $\Upsilon(4S)$ resonance is used for the first decay-time-dependent analysis of these decays within the experiment.
We reconstruct  about 135 signal candidates, and measure $\mathcal{B}(\Bz \to \Kz \piz) = [11.0 \pm 1.2 \stat \pm 1.0 \syst] \times 10^{-6}$ and $\ACP (\Bz \to \Kz \piz)= -0.41_{-0.32}^{+0.30}\stat \pm 0.09\syst$.
 
\end{abstract}


\maketitle

{\renewcommand{\thefootnote}{\fnsymbol{footnote}}}
\setcounter{footnote}{0}




\section{Introduction}
The $\Bz\to\Kz\piz$ decay is mediated by flavor-changing neutral currents.
In the standard model (SM), the dominant decay amplitude is given by the $b\to sd\dbar$ loop, which is dominated by the top quark contribution and carries a weak phase arg$\left(V_{tb}V_{ts}^{*}\right)$.
Here, $V_{ij}$ denote the CKM matrix elements.
Such processes are suppressed in the SM and provide an indirect route to search for beyond-the-SM particles that might be exchanged in the loop.
In the $\Bz\to\Kz\piz$ decay, $\CP$ violation can occur either directly in the decay amplitude ($\ACP$) or via the interference between decays with and without $\Bz$--$\Bzb$ mixing ($\SCP$).
Neglecting subleading contributions to the amplitude, $\SCP$ is expected to be equal to $\sin 2\phi_{1}$ and $\ACP\approx 0$, where $\phi_{1}\equiv$ arg$\left(-V_{cd}V^{*}_{cb}/V_{td}V^{*}_{tb}\right)$.
Deviations from these expectations could be due to larger-than-expected subleading SM contributions or from non-SM physics.

Combining $B$-meson lifetimes ($\tau$) with branching fractions ($\mathcal{B}$) and direct $\CP$ asymmetries of four $B\to K\pi$ decays related by isospin symmetry, the sum rule proposed in Ref.~\cite{Gronau}, 
\begin{eqnarray}
 \ACP({K^+\pi^-}) + \ACP({K^0\pi^+})\frac{\mathcal{B}(K^0\pi^+)}{\mathcal{B}(K^+\pi^-)}\frac{\tau_{B^0}}{\tau_{B^+}}\\ \nonumber
 - 2\ACP({K^+\pi^0})\frac{\mathcal{B}(K^+\pi^0)}{\mathcal{B}(K^+\pi^-)}\frac{\tau_{B^0}}{\tau_{B^+}} - 2\ACP({K^0\pi^0})\frac{\mathcal{B}(K^0\pi^0)}{\mathcal{B}(K^+\pi^-)}
 =0,
\end{eqnarray}
is expected to hold with an uncertainty below $1\%$ and provides an important consistency test of the SM.
Deviations from this isospin sum rule can be caused by an enhancement of color-suppressed tree amplitudes, or by contributions from non-SM physics.
The  prediction of the $\CP$ asymmetry $\ACP (\Kz \piz)$ from this sum-rule is $-0.138 \pm 0.025$~\cite{kpisensitivity}, using up-to-date known values of other quantities~\cite{HFLAV}.
Combining measurements from Belle and BaBar~\cite{belle,babar}, the Heavy Flavor Averaging Group finds $\ACP=0.01 \pm 0.10$ ~\cite{HFLAV}.
The dominant contribution to the uncertainty in this sum-rule comes from the uncertainty in $\ACP({K^0\pi^0})$.
Therefore, a precise measurement of $\ACP({K^0\pi^0})$ is crucial for this consistency test of the SM.

Preliminary results on $\mathcal{B}$ and $\ACP$ of $B^{0}\to \Kz\pi^{0}$ decays have been reported by Belle~II using a data sample corresponding to $62.8\invfb$.
In this analysis, we utilize a larger data set ($189.8\invfb$) and further enhance our sensitivity to $\ACP$ by using $B$ decay-time information.

At Belle~II, pairs of neutral $B$ mesons are coherently produced in the process $e^{+}e^{-}\to \Upsilon(4S)\to \Bz\Bzb$.
When one of the $B$ mesons decays to a $\CP$ eigenstate $f_{\CP}$, such as $\KS\piz$, and the other to a flavor-specific final state $f_{\rm tag}$, the time-dependent decay rate is given by
\begin{eqnarray}
\mathcal{P}(\Delta t)=\frac{{\rm e}^{-|\Delta t|/\tau_{\Bz}}}{4\tau_{\Bz}}[1+q\{\ACP\cos(\Delta m_{d}\Delta t)+\SCP\sin(\Delta m_{d}\Delta t)\}],
\label{equation:eqn1}
\end{eqnarray}
where $\Delta t=t_{\CP}-t_{\rm tag}$ is the proper-time difference between the decays into $f_{\CP}$ and $f_{\rm tag}$, $q$ equals $+1$ ($-1$) for the $\Bz$ ($\Bzb$) decay to $f_{\rm tag}$, and $\Delta m_{d}$ is the $\Bz$--$\Bzb$ mixing frequency.
This analysis employs a decay-time-dependent $\CP$ asymmetry fit similar to the previous measurement of $\sin 2\phi_{1}$~\cite{tsi}.
The key  challenge here lies in the determination of the position of the $\Bz\to\Kz\piz$ decay vertex.
For that, the $\KS$ flight direction is projected back to the interaction region and the $\KS$ is required to decay inside the vertex detector (VXD).
The full analysis was developed and tested with simulated data, and validated with data control samples before selecting and inspecting the $\Bz\to\Kz\piz$ candidates.
Due to the limited sensitivity provided by the available data sample, we measure $\ACP$ by fixing $\SCP$, $\Delta m_{d}$, and $\tau_{\Bz}$ to their known values~\cite{HFLAV}.

\section{The Belle II detector and data sample}

Belle II~\cite{belle2det} is a particle spectrometer having almost $4\pi$ solid-angle coverage, designed to reconstruct final-state particles of $\ep\en$ collisions delivered by the SuperKEKB asymmetric-energy collider~\cite{supkek}. It is located at the KEK laboratory in Tsukuba, Japan.
The energies of the positron and electron beams are $4$ and $7\gev$, respectively.
Belle II consists of a number of subdetectors surrounding the interaction region in a cylindrical geometry.
The innermost one is the VXD, comprised of several position-sensitive silicon sensors. 
It samples the trajectories of charged particles (`tracks') in the vicinity of the interaction region to determine the decay positions of their  parent particles.
The VXD includes two inner layers of  pixel sensors and four outer layers of double-sided microstrip sensors.
The second pixel layer is currently incomplete covering  one sixth of the azimuthal angle.
Charged-particle momenta and charges are measured by a large-radius, small-cell, central drift chamber (CDC), which also offers particle-identification information via a measurement of specific ionization.
A  Cherenkov-light angle and time-of-propagation detector surrounding the CDC provides charged-particle identification in the central detector volume, supplemented by proximity-focusing, aerogel, ring-imaging Cherenkov detectors in the forward region with respect to the electron beam.
A CsI(Tl)-crystal electromagnetic calorimeter (ECL) provides energy measurements of electrons and photons. 
A solenoid surrounding the ECL generates a uniform axial 1.5\,T magnetic field.
Layers of plastic scintillators and resistive-plate chambers, interspersed between the magnetic flux-return iron plates, allow for the identification of $\KL$ mesons and muons.
The subdetectors most relevant for our study are the VXD, CDC, and ECL.

We analyse collision data collected at a center-of-mass (CM) energy near the $\Upsilon (4S)$ resonance, corresponding to an integrated luminosity of $189.8\invfb$.
We use large samples of simulated $e^{+}e^{-}\to \qqbar$ $(q=u,d,s,c)$, $\Upsilon(4S)\to\BzBzb$ and $\BpBm$ events to optimize the event selection and study possible background contributions.
Simulated $\Bz \to \KS \piz$ signal events are used to determine signal models and estimate the selection efficiency.
We use the \textsc{EVTGEN} package~\cite{evtgen} to generate $B$-mesons decays and the \textsc{PHOTOS} package~\cite{photos} to calculate final-state radiation from all charged particles.
The simulation of $e^+e^- \to \qqbar$ continuum background relies on the \textsc{KKMC} generator~\cite{kkmc} interfaced to \textsc{Pythia}~\cite{pythia}.
The interactions of final-state particles with the detector are simulated using \textsc{Geant4}~\cite{geant}.

\section{Reconstruction and selection}

Tracks are reconstructed with the VXD and CDC.
Photons are identified as isolated energy clusters in the ECL that are not matched to any track.
Candidate \KS mesons  are reconstructed from pairs of oppositely-charged particles with the dipion mass  between $482$ and $513\mevcc$.
We reconstruct \piz candidates from pairs of photons that have energies greater than 80 (223)\,\mev if detected in the barrel (endcap) ECL.
We apply the different energy thresholds to suppress beam background, which is higher in the endcap compared to the barrel region.
The selection also requires the diphoton mass to lie between $119$ and $150\mevcc$ and the absolute value of the cosine of the angle between each photon and the $B$ meson in the $\piz$ rest frame to be less than 0.953.
These criteria suppress contributions from misreconstructed $\piz$ candidates. 

A $B$-meson candidate is reconstructed by combining a $\KS$ with a $\piz$ candidate.
For this purpose, we use two kinematic variables, the beam-energy-constrained mass $(M_{\rm bc})$ and the energy difference $(\Delta E)$,
\begin{align}
M_{\rm bc} &= \sqrt{E_{\rm beam}^{2} - \vec{p}_{B}^{\,2}},\\ \nonumber
\Delta E &= E_{B} - E_{\rm beam},
\end{align}
where $E_{\rm beam}$ is the beam energy, and $E_{B}$ and $\vec{p}_{B}$ are respectively the reconstructed energy and momentum of the $B$ meson; all calculated in the CM frame.
 
The presence of a high momentum $\piz$ causes a significant correlation between $M_{\rm bc}$ and $\Delta E$ due to the shower leakage of final-state photons.
To reduce this correlation, we use a modified version of $M_{\rm bc}$ that is defined in terms of the beam energy and momenta of final-state particles as
\begin{eqnarray}
M^{\prime}_{\rm bc}=\sqrt{E_{\rm beam}^{2}-\left( \vec{p}_{\KS} + \frac{\vec{p}_{\piz}}{|\vec{p}_{\piz}|} \sqrt{(E_{\rm beam} - E_{\KS})^{2}- m_{\piz}^{2}} \right)^{2}},
\end{eqnarray}
where all kinematic quantities are again calculated in the CM frame.
We retain candidate events satisfying $ 5.24<M^{\prime}_{\rm bc}<5.29~\gevcc$ and $|\Delta E|<0.30~\gev$. 

To measure the proper-time difference $\Delta t$, we need to determine the signal and tag-side $B$ decay vertices.
The signal $B$ vertex is obtained by projecting the flight direction of the $\KS$ candidate back to the interaction region.
The $\KS$ flight direction is determined from its decay vertex and momentum.
The intersection of the $\KS$-flight projection with the interaction region provides a good approximation of the signal $B$ decay vertex, since both the transverse flight length of the $\Bz$ meson and the transverse size of the interaction region are small compared to the $\Bz$ flight length along the boost direction.
The tag-side vertex is obtained with tracks that are not associated to the $\Bz\to\KS\piz$ decay.
We obtain  $\Delta t$  by dividing the longitudinal distance between the signal and tag vertices by the speed of light and the Lorentz boost of the $\FourS$ system in the lab frame.
Signal candidates with poorly measured $\Delta t$, mainly due to $K^{0}_{S}$ mesons decaying outside of the VXD acceptance, are suppressed by requiring the estimated uncertainty on $\Delta t$ to be less than 2.5\,ps.

\begin{figure}[htb!]
    \centering
    \includegraphics[scale=0.45]{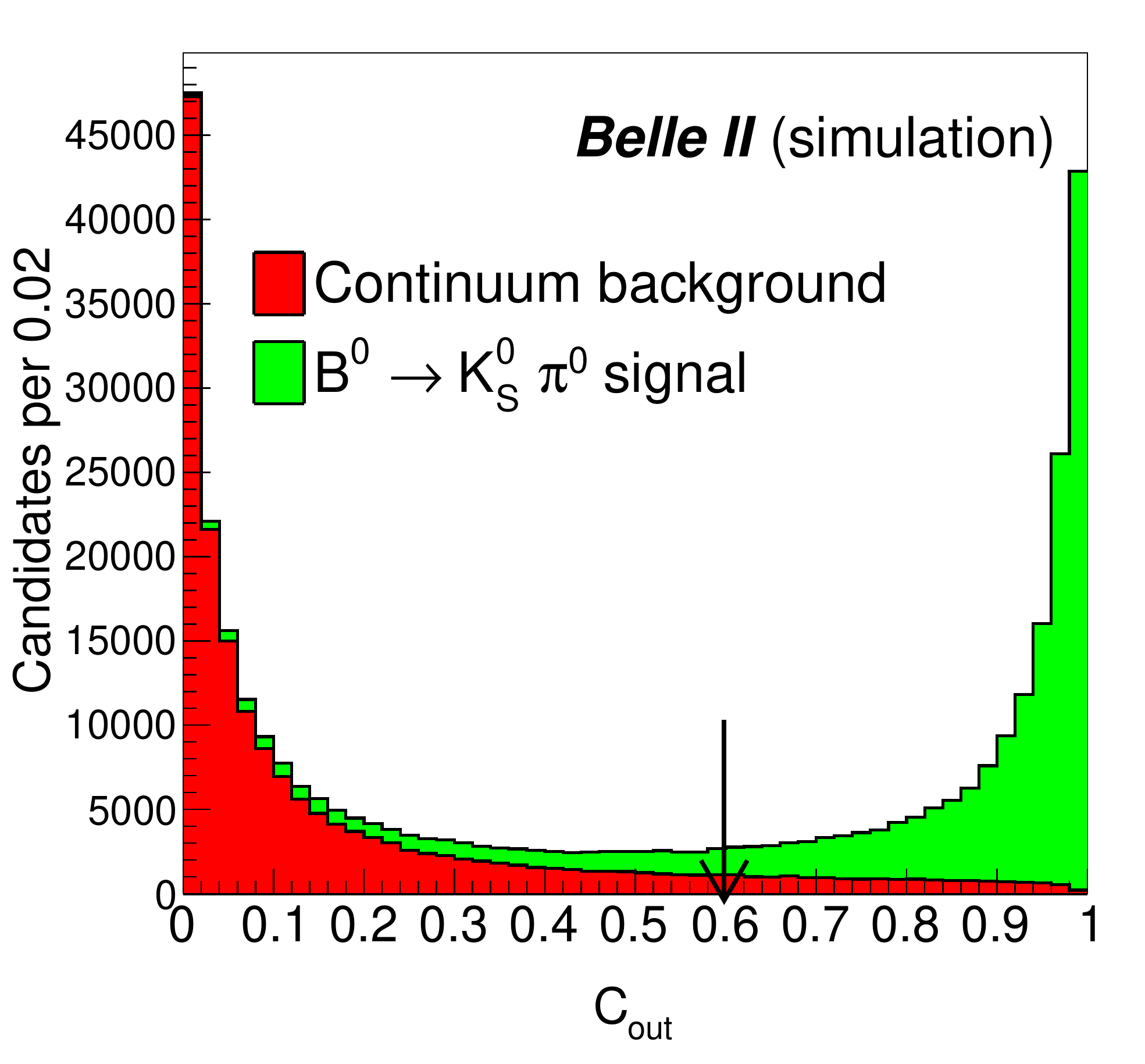}
       \caption{Distributions of the BDT output $C_{\rm out}$ for simulated signal and $e^{+}e^{-}\to q\qbar$ events. The downward arrow indicates the position of the applied $C_{\rm out}$ selection.}
    \label{fig:bdt}
\end{figure}

Events from continuum $e^{+}e^{-}\to \qqbar$ production are suppressed using a boosted-decision-tree (BDT) classifier~\cite{bdt} that exploits several event-topology variables known to provide discrimination between $B$-meson signal and continuum background.
The following variables are those offering most of the discrimination:
modified Fox--Wolfram moments~\cite{ksfw}, CLEO cones~\cite{cleo}, the magnitude of the thrust axis for the reconstructed $B$ candidate, and the cosine of the angle between the thrust axis of the signal $B$ and that of rest of event.
The BDT is trained on samples of simulated $e^{+}e^{-}\to q\qbar$, $\Bz\Bzb$ and $B^{+}B^{-}$ events, each equivalent to an integrated luminosity of $1\invab$.
The BDT output distribution ($C_{\rm out}$) is shown in Fig.~\ref{fig:bdt}.
We apply a criterion $C_{\rm out}>0.60$, which rejects about $89\%$ of the continuum background with a $18\%$ relative loss in signal efficiency.
We then translate $C_{\rm out}$ into a new variable,
\begin{equation}
C_{\rm out}^{\prime}= {\rm ln} \left(\frac{C_{\rm out}-C_{\rm out,min}}{C_{\rm out,max}- C_{\rm out}}\right),
\end{equation}
where $C_{\rm out,min}=0.60$ and $C_{\rm out,max} = 0.99$.
The distributions of $C_{\rm out}^{\prime}$ can be  parametrized with Gaussian functions.

After applying all selection criteria, the average number of $B$ candidates per event is 1.009.
Multiple candidates arise due to random combinations of final-state particles.
To select the best combination in an event with multiple candidates, we first compare the $\piz$ mass-constrained fit $ \chi^{2}$ probability (`p-value').
If there are two or more $B$ candidates sharing the same $\piz$, we choose the one with the best p-value of the fit of the $\KS$ vertex.
This selection retains the correct $B$ candidate in $74\%$ of simulated signal events.

The signal efficiency ($\epsilon$) of correctly reconstructed  events after all selection criteria have been applied is $12.3\%$.
From simulation we find that signal candidates can be incorrectly reconstructed in $1.5\%$ of the times by accidentally picking up a particle from the other $B$ meson decay.

We determine the flavor of the tag-side $B$ meson ($q$) from the properties of the final-state particles that are not associated with the reconstructed $\Bz\to\KS\piz$ decay.
The Belle~II multivariate flavor-tagger algorithm~\cite{flavortagger} uses the information of $B$-decay products to determine the quark-flavor of $B$ mesons.
It gives two parameters, the $b$-flavor charge, $q$ and its quality factor $r$.
The parameter $r$ is an event-by-event, MC determined flavour-tagging dilution factor
that ranges from $0$ (no flavor discrimination) to $1$ (unambiguous flavor assignment).
\section{Determination of branching fraction and $\CP$ asymmetry }
\label{sec:yields}

We obtain the signal yield and direct $\CP$ asymmetry from an extended maximum-likelihood fit  to the unbinned distributions of $M^{\prime}_{\rm bc}$, $\Delta E$, $C_{\rm out}^{\prime}$, and $~\Delta t$.
For the signal component, $M^{\prime}_{\rm bc}$ is modeled with the sum of a Crystal Ball ~\cite{CB} and a Gaussian function with a common mean; $\Delta E $ with the sum of a Crystal Ball and two Gaussian functions, all three with a common mean; and $C_{\rm out}^{\prime}$  with the sum of an asymmetric and a regular Gaussian function.
The signal $\Delta t$ probability density function (PDF) is given by
\begin{eqnarray}
\label{equation:tdcpv}
\mathcal{P}_{\rm sig}(\Delta t, q) &= &\frac{{\rm e}^{-|\Delta t|/ \tau_{\Bz}}}{4 \tau_{\Bz}} [\{1-q \Delta w_{r} + q \mu_{r}(1-2w_{r})\}+\{q(1-2w_{r})+\mu_{r}(1-q \Delta w_{r})\}\\ \nonumber
& & \{\ACP \cos(\Delta m_{d} \Delta t)+
\SCP \sin(\Delta m_{d} \Delta t)\}] \otimes \mathcal{R}_{\rm sig},
\end{eqnarray}
where $w_{r}$ is the fraction of incorrectly tagged events, $\Delta w_{r}$ is the difference in $w_{r}$ between $\Bz$ and $\Bzb$, $\mu_{r}$ is the difference in their tagging efficiency (that is the fraction of signal $\Bz$ or $\Bzb$ candidates to which a flavor tag can be assigned), and $\mathcal{R}_{\rm sig}$ is the $\Delta t$ resolution function.
The function $\mathcal{R}_{\rm sig}$ is composed of a sum of two Gaussians with a combined width of $\approx$ 0.9\,ps, and its parameters are determined with simulated events.
We set $\tau_{\Bz}$ to 1.520\,ps, $\Delta m_{d}$ to 0.507\,$\rm{ps}^{-1}$, and $\SCP$ to 0.57~\cite{HFLAV}. 
The data are divided into seven $q\times r$ bins with the tagging parameters for each bin ($w_{r}$, $\Delta w_{r}$, and $\mu_{r}$) fixed to the corresponding values~\cite{flavortagger}.
The effective tagging efficiency $\epsilon_{\rm eff}$ ($=\sum_{r} \epsilon_{r}\times (1- 2 w_{r})^{2}$, where $\epsilon_{r}$ is the partial effective efficiency in the $r$-th bin), $w_{r}$, and $\mu_{r}$ are $(30.0\pm 1.2) \%$, $(2$--$47)\%$, and $(0.5$--$11)\%$, respectively.
All signal PDF shapes are fixed to the values determined from a $q\times r$ binned fit to simulated events.

For the continuum background component, an ARGUS function~\cite{AG} is used for $M^{\prime}_{\rm bc}$, a linear function for $\Delta E$, and the sum of an asymmetric and a regular Gaussian function for $C_{\rm out}^{\prime}$.
Its $\Delta t$ distribution is modeled with an exponential function convolved with a Gaussian for the tail; we use a double Gaussian for its resolution function ($\mathcal{R}_{\qqbar}$).
For the continuum background component, we float the PDF shape parameters, which are found to be common for all $q\times r$ bins.
For the $\BB$ background component, a two-dimensional Kernel estimation PDF~\cite{2D} is used to model the $\Delta E$ vs.\ $M^{\prime}_{\rm bc}$ distribution, and the sum of an asymmetric and a regular Gaussian function is used for $C_{\rm out}^{\prime}$.
Its $\Delta t$ distribution is modeled with an exponential function convolved with a Gaussian for the tail; we again use a double Gaussian for its resolution function ($\mathcal{R}_{\BB}$).
The $\BB$ background shape parameters are fixed from a fit to the corresponding simulated sample.

The fit parameters are the signal yield $N_\mathrm{sig}$; $\ACP$; $\BB$ background yield, which is Gaussian constrained to the result of a fit to the $\Delta E$ sideband in data; continuum background yield; $M^{\prime}_{\rm bc}$ ARGUS parameter; $\Delta E$ slope; and effective width of $C^{\prime}_{\rm out}$ for the $\qqbar$ component.
We correct the signal $M^{\prime}_{\rm bc}$, $\Delta E$, and $C_{\rm out}^{\prime}$ PDF shapes for possible data--simulation differences, according to the values obtained with a control sample of $B^{+} \to \Dzb (\to K^{+}\pi^{-} \pi^{0} )\pi^{+}$ (charge conjugated modes are implicitly
included hereafter).
In order to mimic the signal decay, we apply a similar $\piz$ selection. We use a maximum-likelihood fit to the unbinned distributions of $M^{\prime}_{\rm bc}$, $\Delta E$, and $C^{\prime}_{\rm out}$, using PDF shapes similar to those employed to describe the signal in data.
We use a  control sample of $\Bz\to\jpsi(\to\mu^{+}\mu^{-})\KS$ decays to validate the time-dependent analysis.
To mimic the signal decay, we do not use the two muons coming from the $\jpsi$ to reconstruct the signal $B$ decay vertex.
We use a maximum-likelihood fit to the unbinned distributions of $\mbc$ and $\Delta t$, using PDF shapes and resolution functions similar to those employed in the fit to the signal in data. 
The $\Bz$ lifetime and $\ACP$ are measured to be $1.59^{+0.09}_{-0.08}$\,ps and $-0.03\pm 0.10$, respectively, which are consistent with their known values~\cite{HFLAV}.
The uncertainties quoted here are statistical only. 
This provides convincing data-driven support for the time-dependent part of the analysis.
The same sample is also used to correct the $\Delta t$ PDF shape parameters for possible data--simulation differences.
The estimator properties (mean and uncertainty) have been studied in both simplified and realistic simulated experiments and found to be as expected.

\begin{figure}[!htb]
		\includegraphics[scale=0.4]{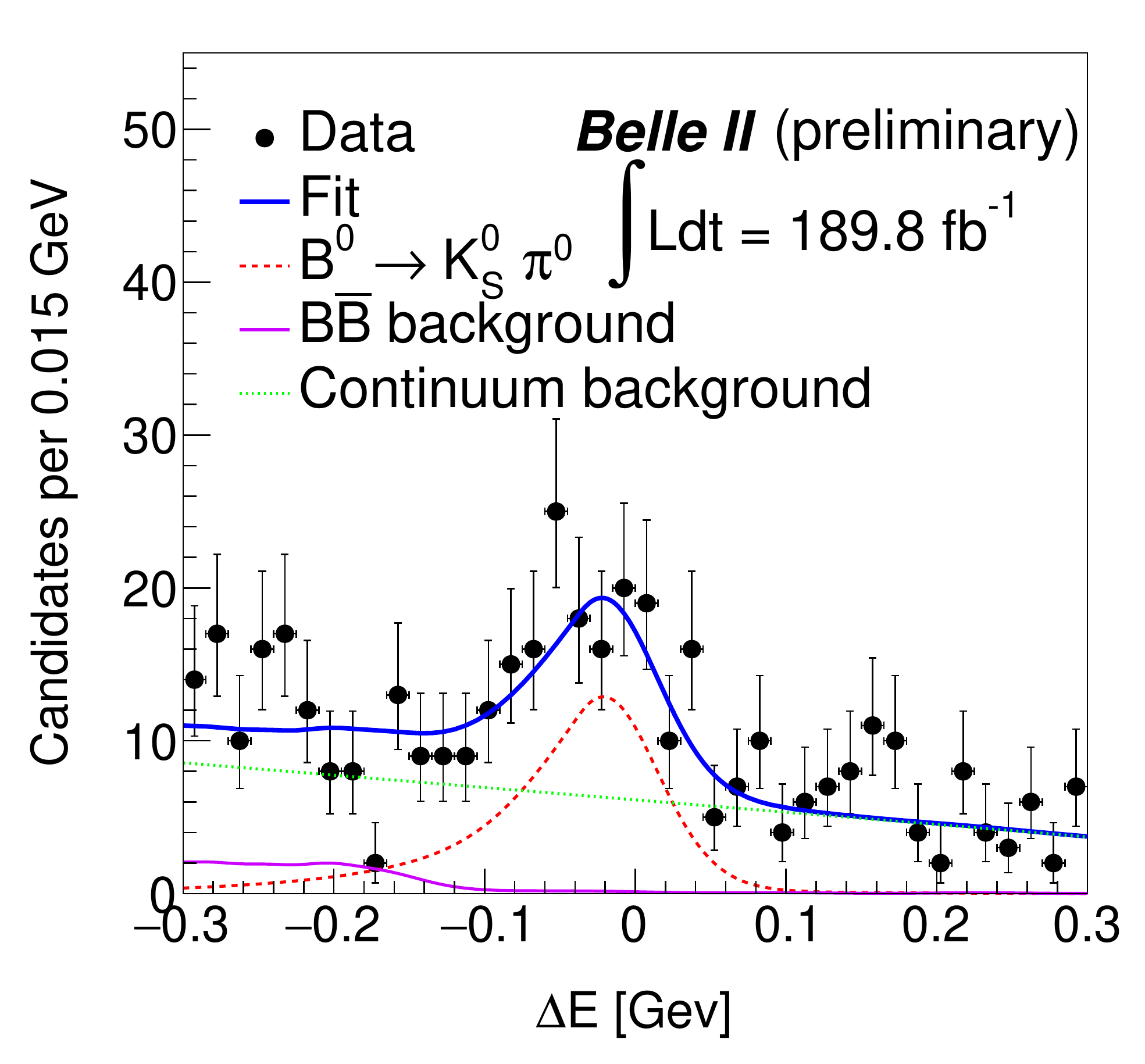}
		\includegraphics[scale=0.4]{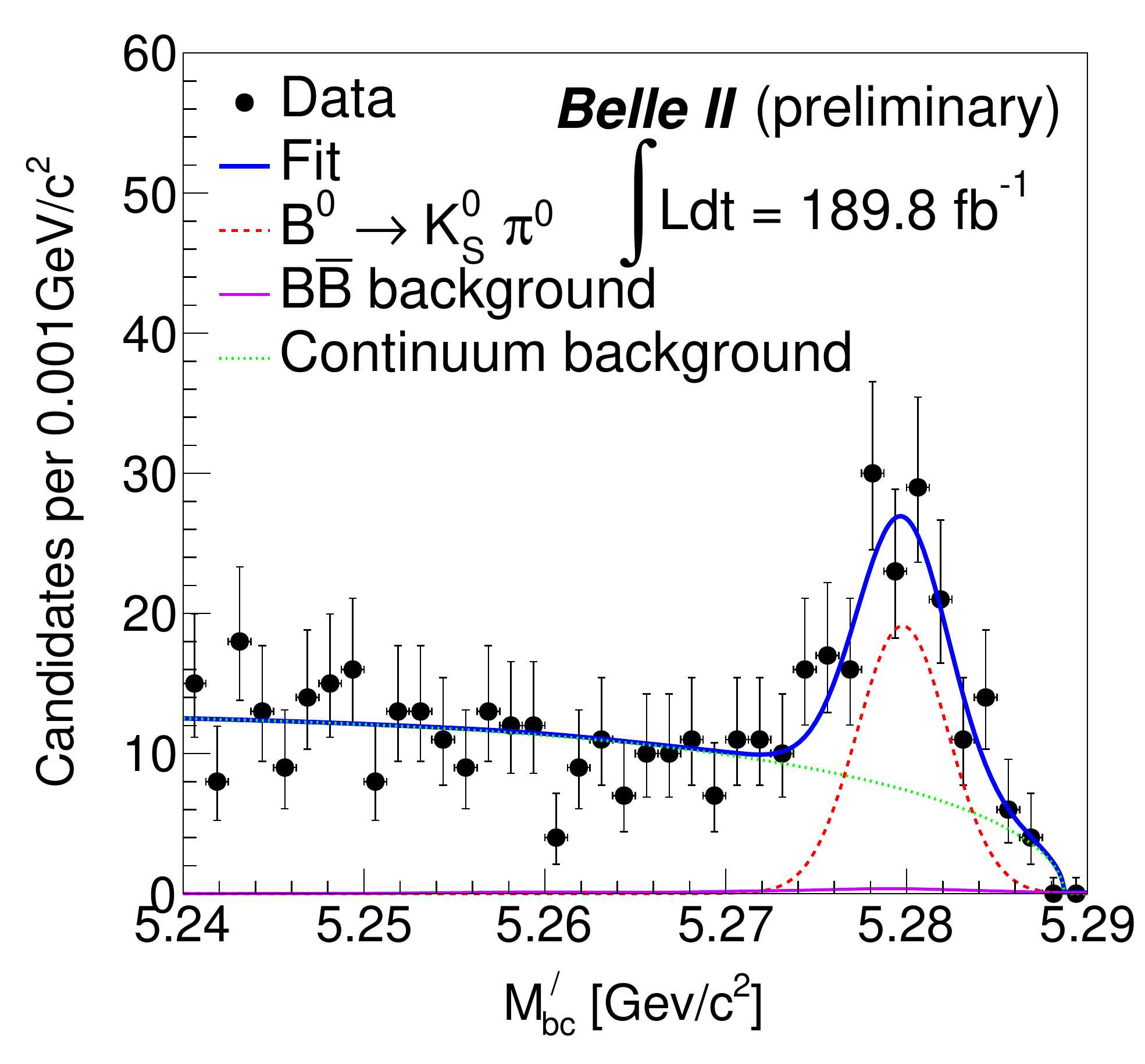} \\
		\includegraphics[scale=0.4]{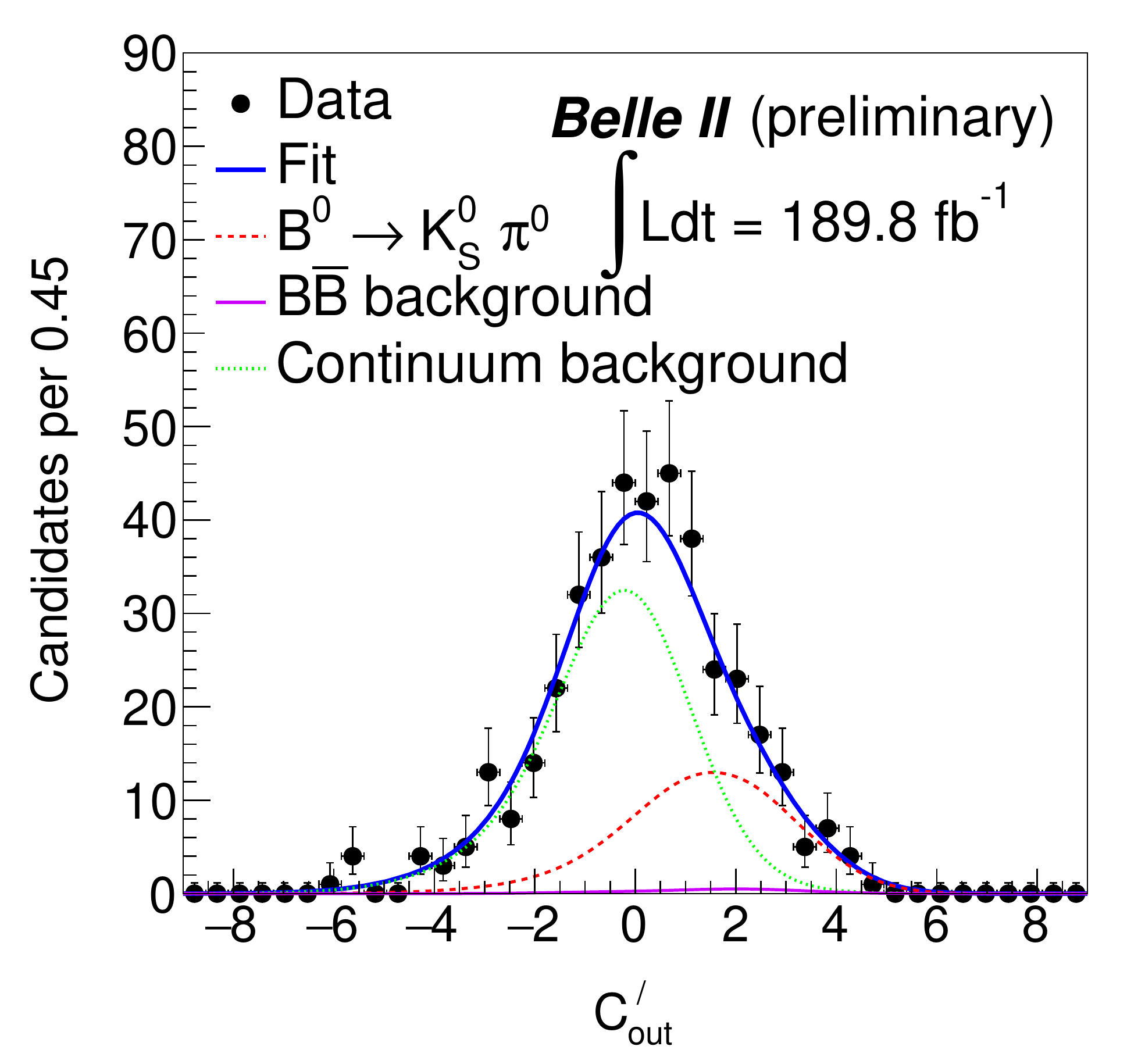} 	
		\includegraphics[scale=0.4]{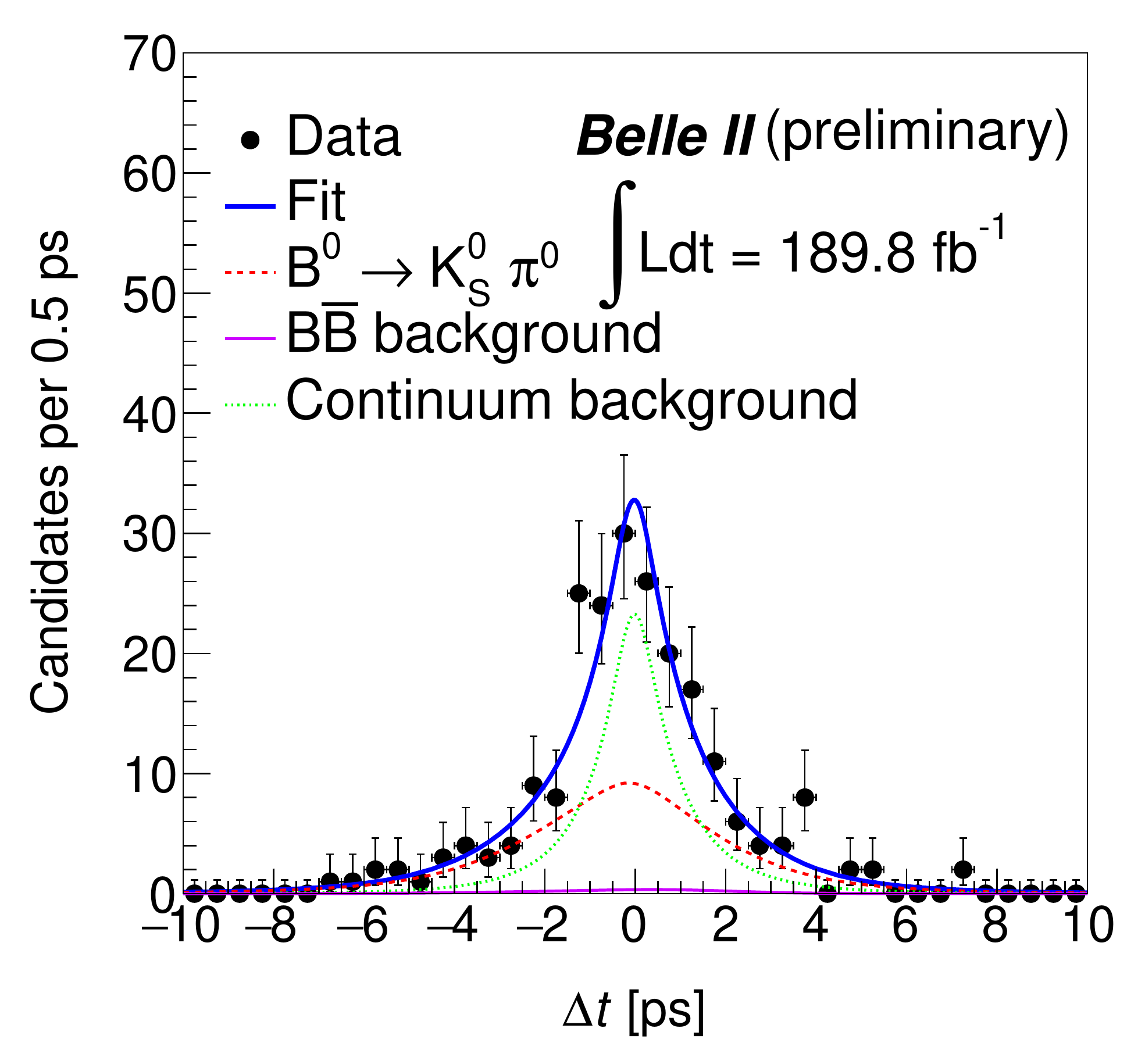} 
		\caption{Signal enhanced fit projections of $\Delta E$ (upper-left), $M^{\prime}_{\rm bc}$ (upper-right), $C^{\prime}_{\rm out}$ (lower-left), and $\Delta t$ (lower-right)  shown for the data sample integrated in the seven $q\times r$ bins.}
	\label{fig:4D}  
\end{figure}

Figure~\ref{fig:4D} shows the four projections of the fit to the seven $q\times r$-integrated data samples which include both $\Bz$  and $\Bzb$ candidates.
For each projection the  signal enhancing criteria, $5.27<M^{\prime}_{\rm bc}< 5.29\gevcc$, $-0.15<\Delta E<0.10\gev$, $|\Delta t|<$ 10.0\,ps, and $C^{\prime}_{\rm out}>0.0$, are applied on all except for the variable displayed.
The obtained signal yield is $135^{+16}_{-15}$, where the quoted uncertainty is statistical only.
We also find $2214^{+49}_{-48}$ continuum and $44\pm 5$ $\BB$ background events.
We determine the branching fraction using the following formula:
\begin{equation}
 \mathcal{B}(\Bz\to\Kz\piz) = \frac{N_{\mathrm{sig}}}{ 2 \times N_{\BB} \times f^{00} \times \epsilon \times {\mathcal{B}}_{s}},  
 \label{equation:branching}
 \end{equation}
 where $N_{\BB}=(197.2\pm 5.70)\times 10^{6}$, $f^{00}=0.487 \pm 0.010$~\cite{f00}, and ${\mathcal{B}}_{s}=0.5$ are the number of $\BB$ pairs, $\FourS \to \Bz\Bzb$ branching fraction, and $\Kz\to\KS$ branching fraction, respectively.
The  $\Bz\to K^{0}\piz$ branching fraction and direct $\CP$ asymmetry ($\ACP$) are measured to be $(11.0 \pm 1.2 \pm 1.0) \times 10^{-6}$ and $-0.41_{-0.32}^{+0.30} \pm 0.09$, respectively.
The first uncertainties are statistical and the second is systematic (described in Section~5).
This extends the previous measurement~\cite{janice} of $\mathcal{B}$ and $\ACP$ in $B^0 \to K^0\pi^0$ decays, where no information on the proper-time difference was used.

\section{Systematic uncertainties}
\label{sec:syst}
The various systematic uncertainties contributing to $\mathcal{B}$ and $\ACP$ are listed in Table~\ref{tab:systematics}.
Assuming these sources to be independent, we add their contributions in quadrature to obtain the total systematic uncertainty.
The systematic uncertainty due to possible differences between data and simulation in the reconstruction of charged particles is $0.3\%$ per track~\cite{tracking}.
We linearly add this uncertainty in $\mathcal{B}$ for each of the two pion tracks coming from the decay of the $\KS$ in the signal $B$.
From a comparison of the $\KS$ yield in data and simulation, we find that the ratio of the $\KS$ reconstruction efficiency changes approximately as a linear function of its flight length~\cite{tracking}.
We apply an uncertainty  of $0.4\%$ for each centimeter of the average flight length of the $\KS$ candidates resulting in a $4.2\%$ total systematic uncertainty in $\mathcal{B}$.
We estimate the systematic uncertainty due to possible differences between data and simulation in the $\pi^{0}$ reconstruction and selection by comparing the inclusive decay sample of $D^{0} \to K^{-} \pi^{+} \pi^{0}$ with $D^{0} \to K^{-} \pi^{+}$~\cite{drate}.
The data--simulation efficiency ratio is found to be close to unity with an uncertainty of $7.5\%$, which we assign as a systematic uncertainty in $\mathcal{B}$.
We evaluate possible data--simulation differences in the continuum-suppression efficiency using the control sample of $B^{+} \to \Dzb (\to K^{+} \pi^{-} \pi^{0} )\pi^{+}$.
As the ratio of efficiencies obtained in data and simulation is close to unity, the statistical uncertainty in the ratio (1.6\%) is assigned as a systematic uncertainty to $\mathcal{B}$.
We estimate the systematic uncertainty in $\ACP$ due to the uncertainty in the  wrong-tag fraction by varying the parameter individually for each $q\times r$ region by its uncertainty.
The systematic uncertainty due to the  $\Delta t$ resolution function is estimated in a similar fashion.
As external inputs $\tau_{B^{0}}$, $\Delta m_{d}$, and $\SCP$ are fixed to their known values in the fit, the associated systematic uncertainties are estimated by varying the values by their uncertainties.
In the nominal fit, we assume the  $\BB$-background decays to be $\CP$ symmetric.
To account for a potential $\CP$ asymmetry in the $\BB$ background, we use an alternative $\Delta t$ PDF given by
\begin{eqnarray}
\label{equation:bbbar_deltat}
\mathcal{P}_{\BB}(\Delta t, q) &= &\frac{{\rm e}^{-|\Delta t|/ \tau_{\Bz}}}{4 \tau_{\Bz}}
[1+q\{\mathcal{A}^{\prime}_{\CP}\cos(\Delta m_{d} \Delta t)+ \mathcal{S}^{\prime}_{\CP}\sin(\Delta m_{d} \Delta t)\}] \otimes R_{\BB}.
\end{eqnarray}
We perform fits to simplified simulated experiments by varying $\mathcal{S}^{\prime}_{\CP}$ and $\mathcal{A}^{\prime}_{\CP}$ from $+1$ to $-1$.
We then calculate the deviations in signal $\ACP$ from its nominal value.
These deviations are assigned as a systematic uncertainty to $\ACP$ due to the asymmetry of the $\BB$ background.
An overall uncertainty of $3.2\%$ in $\mathcal{B}$ is taken as a systematic uncertainty due to the number of $\BB$ pairs used, which also includes the uncertainty in $f^{00}$.
The uncertainties due to the signal PDF shape parameters are estimated by varying their uncertainties. 
Similarly, the uncertainties due to the background PDF shape are calculated by varying all fixed parameters by their uncertainties, determined from the fit to simulated samples.
We fix the $M^{\prime}_{\rm bc}$ ARGUS endpoint to the value obtained from a fit to the $\Delta E$ sideband data.
Subsequently we vary it by $\pm 1\sigma$ to assign a systematic uncertainty, where $\sigma$ is the uncertainty from the fit.
A potential fit bias is checked by performing an ensemble test comprising $1000$ simplified simulated experiments in which signal events are drawn from the corresponding simulation sample and background events are generated according to their PDF shapes.
We calculate the mean shift of the signal yield from the input value and assign it as a systematic uncertainty.
Tag-side interference can arise due to the presence of both CKM-favored and -suppressed tree amplitudes.
The systematic uncertainty in $\ACP$ assigned to this interference is taken from Ref.~\cite{tsi}.
A possible systematic uncertainty related to VXD misalignment is neglected in this study.

\begin{table}[h]
\caption{List of systematic uncertainties contributing to the branching fraction and direct $\CP$ asymmetry.}
\label{tab:systematics}
    \centering
    \begin{tabular}{l|c|c}
    \hline \hline
  Source  & $\delta\mathcal{B}$ (\%) & $\delta\ACP$  \\
 \hline
  Tracking efficiency  & 0.6  & --\\
  $\KS$ reconstruction efficiency & 4.2 & --\\
  $\pi^{0}$ reconstruction efficiency & 7.5 & --\\
  Continuum suppression efficiency &  1.6 & --\\
  Number of $\BB$ pairs &  3.2  & --\\
  Flavor tagging & -- & 0.040  \\
  Resolution function & -- & 0.050 \\
  External inputs&  0.4 &0.021    \\
  $\BB$ background asymmetry & -- & 0.002 \\
  Signal modelling & 1.0 & 0.015 \\
  Background modelling &0.9 & 0.004  \\
  Possible fit bias & 2.0 & 0.010\\
  Tag-side interference & -- & 0.038 \\
  \hline \hline
  Total &  9.6  &0.086\\
    \end{tabular}
\end{table}

\section{Summary}
\label{sec:summary}

We report measurements of the branching fraction and direct $\CP$ asymmetry in $\Bz \to \Kz \piz$ decays using a data sample, corresponding to $189.8 \invfb$ of integrated luminosity, recorded by Belle~II at the $\Upsilon(4S)$ resonance.
The observed signal yield is $135_{-15}^{+16}$.
We measure $\mathcal{B}(\Bz \to \Kz \piz) = [11.0 \pm 1.2\stat \pm 1.0\syst] \times 10^{-6}$ and $\ACP = -0.41_{-0.32}^{+0.30}\stat \pm 0.09 \syst$.
This is the first measurement of $\ACP$ in $\Bz \to \Kz\piz$ performed at Belle II using a decay-time-dependent analysis.
The results agree with previous determinations~\cite{janice,HFLAV}. 
\section{Acknowledgement}
We thank the SuperKEKB group for the excellent operation of the
accelerator; the KEK cryogenics group for the efficient
operation of the solenoid; and the KEK computer group for
on-site computing support.

\end{document}